\documentclass[utf8]{frontiersinFPHY_FAMS-mod} 

\setcitestyle{square} 
\usepackage{url,lineno,microtype,caption}
\usepackage[colorlinks=true]{hyperref}
\usepackage[onehalfspacing]{setspace}
\usepackage{comment}
\def\keyFont{\fontsize{12}{11}\helveticabold }
\def\editFont{\fontsize{14}{12}\helvetica }
\def\firstAuthorLast{Mitsou {et~al.}} 
\def\Authors{Vasiliki A.\ Mitsou\,$^{1}$, Andreas Redelbach\,$^{2*}$,  and Eleni Vardoulaki\,$^{3,4}$}

\hypersetup{
  linkcolor=darkred,   
  citecolor=darkgreen,   
  urlcolor=blue,     
  citebordercolor=darkgreen,
  filebordercolor=0 .5 .5,
  linkbordercolor=1 0 0,
  menubordercolor=1 0 0,
  urlbordercolor=blue,
  runbordercolor=0 .7 .7
}

\def\Address{
$^{1}$Instituto de F{\'i}sica Corpuscular (IFIC), CSIC -- Universitat de Val{\`e}ncia, Paterna, Spain \\
$^{2}$Frankfurt Institute for Advanced Studies and Goethe-Universität Frankfurt, Frankfurt, Germany \\
$^{3}$IAASARS, National Observatory Athens, Athens, Greece \\
$^{4}$Th\"{u}ringer Landessternwarte, Tautenburg, Germany}
\def\corrAuthor{A.\ Redelbach}

\def\corrEmail{\href{mailto:redelbach@fias.uni-frankfurt.de}{redelbach@fias.uni-frankfurt.de}}

\begin{document}
\onecolumn
\firstpage{1}

\title {Promoting Green Computing in High Energy Physics and Astrophysics} 

\author[\firstAuthorLast ]{\Authors} 
\address{} 
\correspondance{} 

\extraAuth{}

\maketitle

\begin{keywords}

\keyFont{\noindent astrophysics, big data, green computing, high energy physics, sustainability} 

\vspace*{0.4cm}
\normalfont

\editFont{\noindent  Editorial on the Research Topic \\
\href{https://www.frontiersin.org/research-topics/62951}{Promoting green computing in high-energy physics and astrophysics}}

\end{keywords}

\section{Introduction}

The field of green computing in high-energy physics and astrophysics is gaining increasing attention due to the exponential growth in data processing demands from experiments in these domains. As these fields continue to expand, the efficient use of computing resources becomes crucial to ensure sustainability and manage the energy consumption associated with large-scale data centers~\cite{Becker:2015,Kern:2015,Lannelongue:2021,Wuttke:2022,Junger:2024,Lannelongue:2023,Lee:2025,Danushi:2025}. Current challenges include optimizing energy needs during the construction, operation, and data analysis phases of experimental facilities. Despite ongoing efforts, there is a pressing need to address the sustainability of computing resources, as these challenges are expected to intensify with future experiments~\cite{Aujoux:2021,Bloom:2025,Boisvert:2025ayi}. Recent studies have highlighted the importance of developing strategies to counteract the rising power consumption trends in data centers. However, there remains a gap in comprehensive approaches that integrate existing technologies and identify areas for further development. This gap underscores the necessity for a detailed overview of current and anticipated energy requirements, which can guide the identification of synergies and collaborative efforts in green computing across high-energy~\cite{Banerjee:2023avd,ATLAS:2025sgg} and astrophysics~\cite{Zwart:2020,Simsek:2024} projects.

This research topic aims to explore and promote strategies for green computing in high-energy physics and astrophysics. The primary objectives include assessing current energy consumption patterns, identifying opportunities for resource optimization, and fostering cross-disciplinary collaborations to enhance sustainability. Key questions to be addressed include: What are the current energy demands in these fields? How can existing technologies be leveraged for more efficient resource usage? What synergies can be identified between high-energy and astrophysics projects to promote green computing?

\section{Overview of contributions}

Sustainable software development is studied in the context of SMASH, a general purpose event generator for heavy-ion collisions~\cite{Sciarra:2024gcz}. Measures to ensure transparency and reproducibility of results are exemplified for this research software in the field of theoretical heavy-ion reactions.  SMASH software includes a standardized set of benchmarks enabling consistent performance comparisons before new releases. Based on the regular benchmarking, one can see that the codebase performance has constantly improved in the last couple of years corresponding to roughly two times acceleration in most of the setups.

A community of researchers with a mindset that includes sustainability aspects when it comes to the use and employment of computing resources can be created through continuous training and education. To this effect, a programme of workshops~\cite{Alimena:2024qpb} was launched within the High Energy Physics department at Deutsches Elektronen-Synchrotron (DESY) in 2023, targeting scientific users of the National Analysis Facility (NAF), hosted at DESY (Hamburg, Germany). To date, it has featured three beginners’ workshops and one advanced workshop, receiving positive participants' feedback. The programme offers user training on best practice across a wide range of applications, such as ROOT, batch computing, git, and continuous integration. An extension of the workshops to include external users of the NAF from outside of DESY is considered as the workshop series continues to grow and mature, such that these tutorials on sustainable computing practices can reach as
many users as possible.

Methods of green computing have been implemented in the ALICE Event Processing Node (EPN) farm for online data reconstruction. As a result of a new computing model and also new computing infrastructure, ALICE can process lead–lead collisions at an interaction rate of up to 50 kHz. Notably, the EPN system performs data reconstruction of all detector systems while data taking is ongoing~\cite{Ronchetti:2024gak}. Outside of beam times, this computing farm can execute asynchronous processing tasks utilizing its 350 high-performance servers, each equipped with four boards for a total of 2,800 GPUs. The decision to use GPUs was imperative due to increased incoming data rates requiring efficient online compression. Coping with detector data from lead-lead collisions at that scale also necessitates efficient online data compression. The developed parallel implementation of entropy coding achieves compression speeds equivalent to 3200 MB/s for 32-bit symbols, a factor of 2 speedup of CPU implementations. Another factor in reducing energy consumption is the modular adiabatic cooling system. Due to the design, adiabatic cooling is more energy-efficient by exploiting the process of evaporation to cool air. Overall energy efficiency of operating the EPN cluster is reflected in a very low value of 1.05 for the power usage effectiveness. 

As the computational requirements of astrophysics and high-energy physics continue to grow, improving the energy efficiency of High Performance Computing (HPC) systems has become essential. This review presents the latest developments in energy-efficient HPC, covering advances in hardware, software, programming models, data management, and application optimization~\cite{suarez25}. The authors discuss the transition towards heterogeneous CPU--GPU architectures, intelligent power management, efficient data movement, and scalable parallel algorithms, highlighting how these technologies can reduce energy consumption while maintaining or improving scientific throughput. The review also emphasizes the role of application developers, encouraging the use of performance analysis, benchmarking and portable software frameworks. FAIR data practices are adopted to maximize both computational performance and the scientific value of large-scale simulations and data processing in astrophysics and particle physics. 

The importance of further optimizing algorithms and computational systems to achieve a balance between performance and energy efficiency, especially in the context of increasing data volumes and heightened demands for computational power, is showcased in a study of a track fitting algorithm. An analysis of the efficiency of the Kalman Filter-based fitting algorithm, which plays a key role in particle trajectory reconstruction, was conducted~\cite{Kozlov:2025jce}. As a testbed, the Compressed Baryonic Matter (CBM) experiment at FAIR (GSI, Germany), among the most significant upcoming projects in heavy-ion physics, was utilised. The performance and energy efficiency of the Kalman fitter on both CPUs and GPUs was evaluated. The GPU implementation demonstrated up to a threefold improvement in energy efficiency, resulting in a proportional reduction in power consumption and associated CO$_2$ emissions during data processing. Therefore, optimising the algorithm for GPU execution not only enhances computational efficiency but also significantly reduces the carbon footprint. The analysis provided a quantitative estimate of the carbon footprint associated with track reconstruction and demonstrates how hardware choices influence overall emissions in large-scale data processing workflows.

A method to compute the energy efficiency for the first high level trigger (HLT1) of the LHCb experiment is presented~\cite{Zhuo:2026xzy}. The HLT1 system performs real time event reconstruction and selection at 30 MHz proton-proton collision rate using approximately 500 GPUs.  The model derived in this study relates the throughput with GPU hardware specifications as a basis to predict also the energy efficiency, i.~e. how many events the application can reconstruct per Joule of energy consumed. The models thus allows evaluating both throughput and energy efficiency together when selecting GPUs for any GPU-based processing system. Interestingly, the results for energy efficiency show that the predicted values agree with the measured values within a few percent for all GPUs investigated. The study was carried out using the HLT1 at LHCb as the benchmark framework, however, the approach of modeling throughput and power consumption as functions of GPU specification parameters can be applied to any GPU-based application where performance needs to be projected across different hardware generations.

\begin{figure}[h!]
\begin{center}
\includegraphics[width=\linewidth]{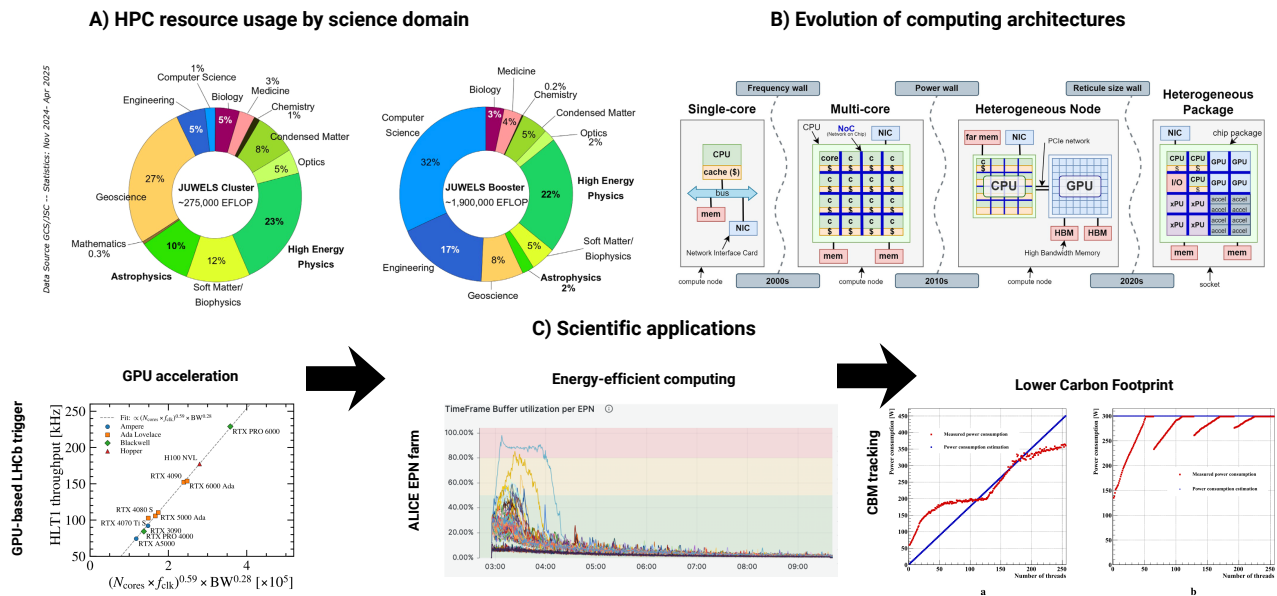}
\end{center}
\caption{\small Overview (thumbnail-sized) of key aspects of sustainable computing in high-energy physics and astrophysics highlighted in the Research Topic. 
Panel~A illustrates the distribution of HPC resource usage across scientific domains, demonstrating the significant demands of high-energy physics and astrophysics \citep[adopted from Fig.~1 of][]{suarez25}. Panel~B shows the evolution of computing architectures from single-core processors to modern heterogeneous CPU and GPU systems, highlighting the scientific progress in energy-efficient strategies \citep[adopted from Fig.~2 of][]{suarez25}. Panel~C presents representative examples of sustainable computing in practice: GPU-acceleration event processing in the LHCb experiment \citep[left, adopted from Fig.~5 of][]{Zhuo:2026xzy}, energy-efficient large-scale computing infrastructure in the ALICE experiment \citep[middle, adopted from Fig. 4-top of][]{Ronchetti:2024gak}, and reduced energy consumption and carbon footprint in the CBM experiment \citep[right, adopted from Fig.~2 of][]{Kozlov:2025jce}. }
\label{fig:1}
\end{figure}

Key aspects of sustainable computing within this Research Topic are summarized in Figure~\ref{fig:1}.

\section{Implications for broader context}

Results included in this Research Topic illustrate how advances in computing architecture enable more energy-efficient computing and also  reduced environmental impact.
Tools to measure resource consumptions exist and their application should be extended on different levels of software development. Also optimization of software during ongoing established projects can have a significant impact towards more sustainable computing.  A practical approach to be considered is running a compute job at a site with less carbon footprint.  Utilization of software solutions available with a focus on high quality and energy efficiency will be beneficial for many projects in fundamental physics. In order to maximize effects for green computing, resource efficiencies of complete workflows throughout the data processing chain should be validated on a regular basis for a transparent documentation of resource usage.

\section*{Author Contributions}

All authors listed have made a substantial, direct and intellectual contribution to the work, and approved it for publication.

\section*{Funding}

The author(s) declared that financial support was received for this work and/or its publication. VAM acknowledges support by the Spanish MCIU/ AEI / 10.13039/501100011033 and the European Union / FEDER through the grants PID2021-122134NB-C21, PID2024-158190NB-C21 and Severo Ochoa CEX2023-001292-S, and by the Generalitat Valenciana (GV) via the Excellence Grant CIPROM/2021/073.
AR is supported by the Bundesministerium für Bildung und Forschung (BMBF) of Germany (grant number 05P21RFFC1) and Helmholtz Research Academy Hesse for FAIR (project IDs 2.1.4.2.2, 2.1.4.2.3), Germany.

\bibliographystyle{JHEP-vaso}

\bibliography{frontiers}

\end{document}